\pdfoutput=1
\documentclass[preprint,aps,superscriptaddress]{revtex4}
\usepackage{graphicx} 
\usepackage{hyperref}

\begin{document}

\title{Substrate-induced band gap opening in epitaxial graphene}

\author{S.Y.~Zhou}
\affiliation{Department of Physics, University of California,
Berkeley, CA 94720, USA} \affiliation{Materials Sciences Division,
Lawrence Berkeley National Laboratory, Berkeley, CA 94720, USA}

\author{G.-H.~Gweon}
\altaffiliation[Present address:]{Department of Physics, University
of California, Santa Cruz, California 95064, USA}
\affiliation{Department of Physics, University of California,
Berkeley, CA 94720, USA}

\author{A.V.~Fedorov}
\affiliation{Advanced Light Source, Lawrence Berkeley National
Laboratory, Berkeley, California 94720, USA}

\author{P.N.~First}
\affiliation{School of Physics, Georgia Institute of Technology,
Atlanta, GA 30332-0430}

\author{W.A.~de~Heer}
\affiliation{School of Physics, Georgia Institute of Technology,
Atlanta, GA 30332-0430}

\author{D.-H.~Lee}
\affiliation{Department of Physics, University of California,
Berkeley, CA 94720, USA}

\author{F. Guinea}
\affiliation{Instituto de Ciencia de Materiales de Madrid, CSIC,
Cantoblanco, E-28049 Madrid, Spain.}

\author{A.H.~Castro~Neto}
\affiliation{Department of Physics, Boston University, 590
Commonwealth Avenue, Boston, MA 02215,USA}

\author{A.~Lanzara}
\affiliation{Department of Physics, University of California,
Berkeley, CA 94720, USA} \affiliation{Materials Sciences Division,
Lawrence Berkeley National Laboratory, Berkeley, CA 94720, USA}

\date{\today}


\maketitle

{\bf Graphene has shown great application potentials as the host material for next generation electronic devices. 
However, despite its intriguing properties,
one of the biggest hurdles for graphene to be useful as an
electronic material is its lacking of an energy gap in the electronic spectra. 
This, for example, prevents the use of graphene in making transistors. Although several proposals have been made to open a gap
in graphene's electronic spectra, they all require complex engineering of the
graphene layer. Here we show that when graphene is epitaxially grown
on the SiC substrate, a gap of $\approx$ 0.26 eV is produced. This
gap decreases as the sample thickness increases and eventually
approaches zero when the number of layers exceeds four. We propose that the
origin of this gap is the breaking of sublattice symmetry owing to the
graphene-substrate interaction. We believe our results highlight a
promising direction for band gap engineering of graphene.}

Graphene, an atomically thin layer of carbon atoms arranged in a honeycomb lattice, has attracted a lot of research interest because of its intriguing physics as well its application potential \cite{NovoselovSci, NovoselovNat, ZhangNat, BergerSci}.  In particular, the extremely high mobility and the easy control of charge carriers by applying a gate voltage have made graphene a promising material for next generation electronics with properties that may exceed those of conventional semiconductors.  In single layer graphene, the unit cell consists of two carbon atoms - the A
and B sublattices (Fig.1a).  The band structure of graphene exhibits two bands intersecting at two inequivalent points K and K$^\prime$ in the
reciprocal space (Fig.1a). Near these points, the electronic dispersion resembles that of relativistic Dirac electrons.  For this reason, K and
K$^\prime$ are commonly referred to as the ``Dirac points''.  As the valence and conduction bands are degenerate at the Dirac points, graphene is a zero gap semiconductor, and how a gap can be induced is crucial for its application in making devices.  There are two ways to lift the degeneracy of the two bands at the Dirac
points. One is to hybridize the electronic states at K and K$^\prime$, which requires breaking of the translational symmetry \cite{MGV06}. The
other is to break the equivalence between the A and B sublattice, which does not require any translation symmetry
breaking \cite{Trauzettel, Dresselhaus, Fertig, Cohen, Kim, Nilsson, Giovannetti}. To induce these perturbations, various graphene super-structures, such as
graphene quantum dots \cite{Trauzettel}, graphene ribbons \cite{Dresselhaus, Fertig, Cohen, Kim}, and devices based on the combination of
single and bilayer graphene regions \cite{Nilsson, McCann, Eli} have been proposed. Here we show that a gap can be induced in a much easier
and reproducible way in epitaxial graphene on a SiC substrate. As we shall discuss, the interaction between the graphene layer and the substrate will break the A and B sublattice symmetry, which opens up a band gap.

\begin{figure}
\includegraphics[width=13.8 cm] {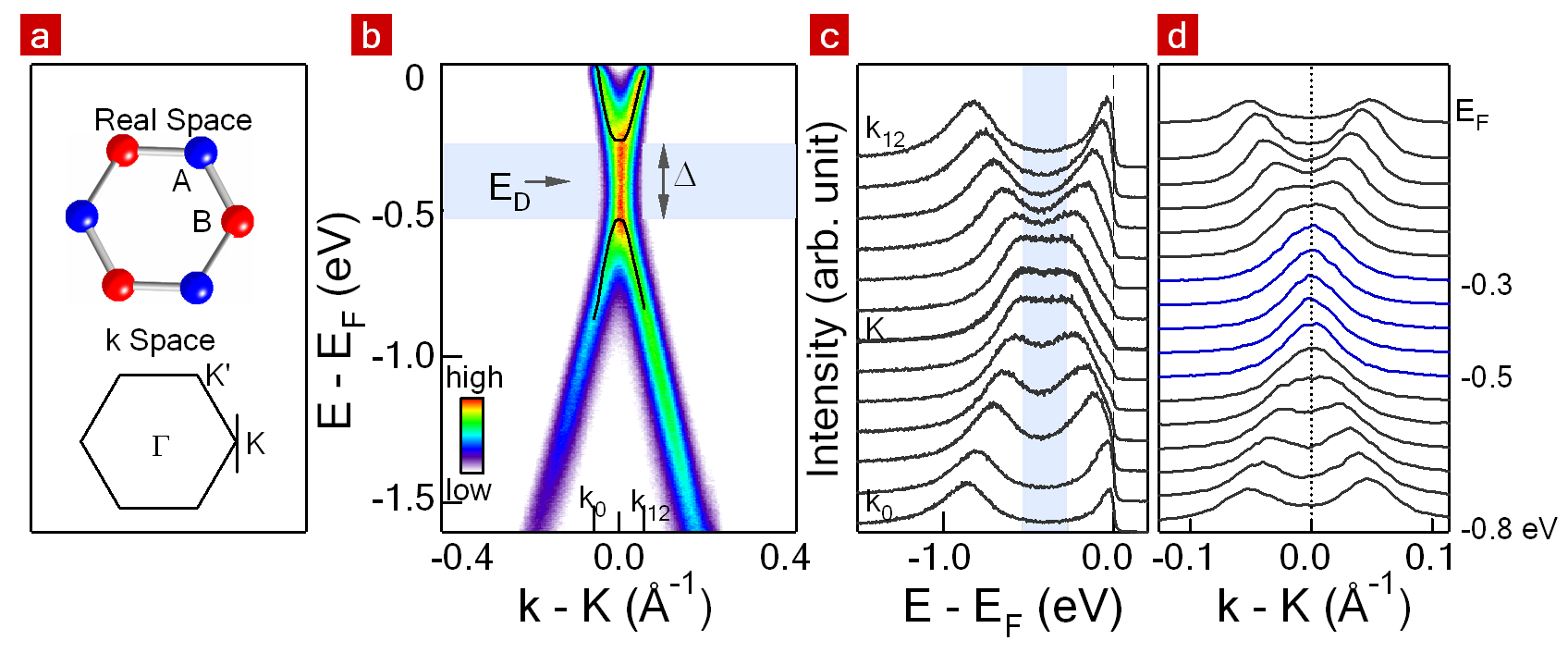}
\label{Figure 1} \caption{ {\bf Observation of the gap opening in single layer graphene at the K point.} (a) Structure of graphene in the real and momentum space.  (b) ARPES intensity map taken along
the black line in the inset of panel (a).  The dispersions (black lines) are extracted from the EDC peak positions shown in panel (c).  (c) EDCs taken near the K
point from k$_0$ to k$_{12}$ as indicated at the bottom of panel
(b).  (d) MDCs from E$_F$ to -0.8 eV.  The blue
lines are inside the gap region, where the peaks are non-dispersive.}
\end{figure}

Figure 1 shows ARPES data taken on single layer graphene for a line through one of the Dirac points, the K point. Panel (b) shows the photoelectron
intensity as a function of energy and the momentum along the black line through K in panel (a).  The black curves
mark the location of peak positions in the energy distribution curves (EDCs).  Following the maximum
in the intensity map, an upward-dispersing and a downward-dispersing cone are clearly observed. This agrees with the expected conical
dispersion of relativistic electrons near the Dirac points.  From the mid-point between the minimum of the conduction band and
the maximum of the valence band, we deduce that, E$_D$, the energy of
the expected Dirac point, is about 0.4 eV below the Fermi energy
(E$_F$). This is in contrast to what is expected for the
undoped graphene where E$_D$=E$_F$, showing that the as-grown graphene is electron doped \cite{Liz, BergerSci}. 
Surprisingly, the dispersion at E$_D$, i.e. the intersection of the cones, is not characterized by a single point as expected for monolayer graphene.  Instead, the valence and conduction bands are separated by a finite energy even at the K point and a gap-like feature is observed. 
This directly follows from the analysis of the EDCs shown in panel (c) and the momentum distribution curves (MDCs) in panel (d). Near the K point, the EDCs show always {\it two peaks} with the minimum energy separation, or the gap, being realized at K.  From this we deduce a gap of $\approx$ 0.26 eV.  The MDC peaks are non-dispersive within the same energy window, 0.26 eV around E$_D$ (blue lines in panel d).  Clearly away from this region, the MDC peaks start dispersing again, in agreement with a conical dispersion.  

A peculiar feature of this gap is that there is non-zero intensity around E$_D$ (panel b) between the valence and conduction bands.   Does this mean that there is no gap at the K point?  If we consider cuts away from the K point (see Fig.S3(a) in supplementary information), based on the conical dispersion of Dirac fermions, a large gap is expected.  However, even this large gap is characterized by a finite intensity.  Therefore we conclude that the non-zero intensity does not mean absence of a gap and in fact there is a gap at the K point.  The remaining question is what is causing this finite intensity inside the gap.  This is very likely the result of the broad EDC peaks (Fig. 1(c)) which cause an overlap of the intensity tails from the top of the valence band and the bottom of the conduction band.  Although at this stage it is not clear why the EDC peaks are so broad, possible causes may be a self energy effect or distribution of gaps.  Finally regardless of the origin, what the large EDC width implies in terms of actual device application remains to be seen in the future.  One should note that ARPES lifetime determined as the inverse line width tends to underestimate the transport lifetime by as much as two orders of magnitude \cite{Calandra}, and that in general one would expect a sharpening up of peaks as they are brought to the Fermi level, as would happen in device applications. 

We note that similar data have been reported recently and discussed in terms of electron-plasmon interaction \cite{EliNatPhys}.  This interpretation is based on the departure of the dispersion from the anticipated behavior near E$_D$ and the observation of an anomalous upturn of the MDC width near E$_D$.  However, these are {\it not} unique features of the K point and they occur every time a gap is present in the spectra.  To discuss the MDC width near the top or bottom of the band in terms of many-body interactions is misleading, as this anomalous upturn of the width often occurs in ARPES near the bottom or the top of a band and is thus an artifact of the MDC analysis.  This is one of the reasons why EDC analysis is more appropriate to extract both the dispersion and the life time in this context.  Finally, since $E_D$ is not at the Fermi energy, within the explanation proposed in Ref.\onlinecite{EliNatPhys} it requires a coincidence for the plasmon feature to center around $E_D$.  In addition, it is quite unlikely that the plasmon energy changes by a factor of two from the single layer to the bilayer graphene, where a similar tail is also observed.

\begin{figure}
\includegraphics[width=16.8 cm] {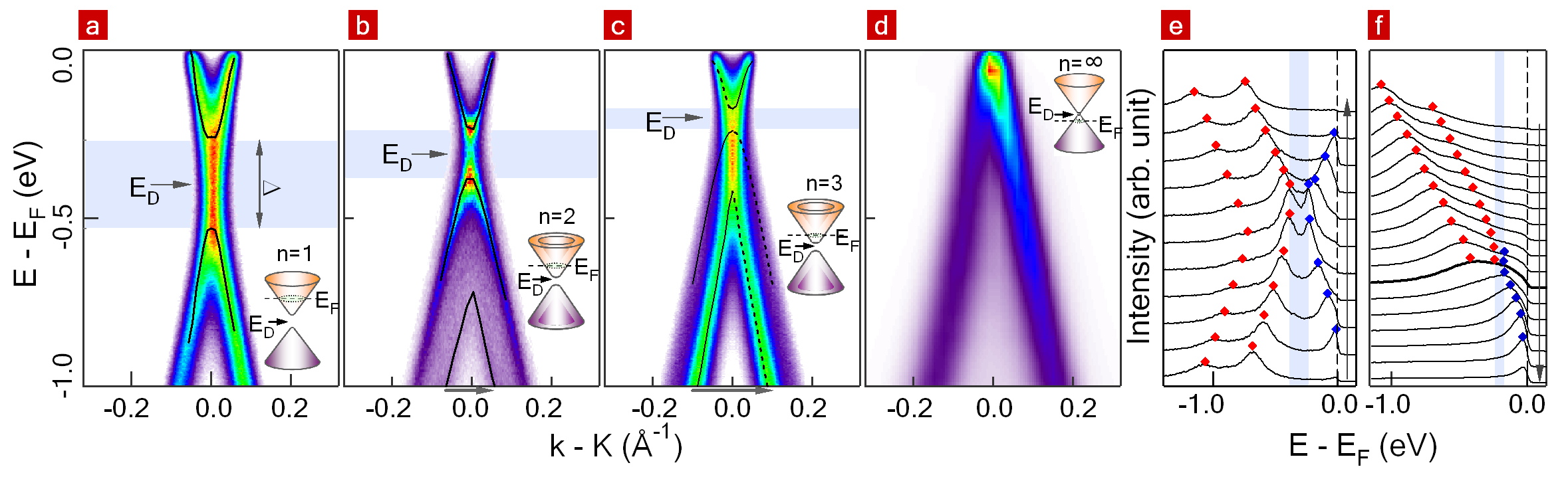}
\label{Figure 2} \caption{ {\bf Decrease of the gap size as the sample becomes thicker.} (a-d) ARPES intensity maps taken on single layer graphene on 6H-SiC, bilayer graphene on 4H-SiC, trilayer graphene on 6H-SiC and graphite respectively.  Data were taken along the black line in the inset of Fig.1(a) except panel (c), which was measured along $\Gamma$K direction and symmetrized with respect to the K point to remove the
strong intensity asymmetry induced by dipole matrix element
\cite{BZselection}.  (e, f) EDCs taken from the raw data (without
symmetrization) for momentum regions labeled by the arrows at the
bottom of panels (b) and (c).}
\end{figure}

Figures 2(a-c) show how the gap and the distance between E$_D$ and
E$_F$ change as the graphene sample thickness varies. Panels (b) and
(c) show the ARPES data for bilayer and trilayer graphene samples.
Again the dispersions extracted from the EDCs (panels (e) and
(f)) are plotted.  In these two panels, two distinct cones can be
identified for E$<$E$_D$.  This is attributed to the splitting of the
$\pi$ bands induced by the interlayer coupling, similar to the
$\approx$ 0.7 eV splitting observed in bulk graphite in the k$_z$=0
plane \cite{NatPhys, AnnalsPhys}.  The absence of the $\pi$ band
splitting in panel (a) and the increase of the splitting from panel
(b) to panel (c) is also a consistent check for the sample thickness
determined by other methods \cite{BergerJPC, Liz}. Panel (d) shows
the ARPES data taken along a line through the H point in graphite, where the
dispersion resembles that of graphene through K \cite{NatPhys}. Data shown in panels (a-d)
 allow us to determine how the electronic structure near
K point varies as the sample thickness increases.  
First of all, as the
sample thickness increases, E$_D$ shifts toward E$_F$. From single layer
to trilayer graphene, E$_D$ (marked by arrows in panels (a-c))
shifts from -0.4 eV to -0.29 eV then to -0.2 eV.  For graphite,
E$_D$ has been estimated to be at $\approx$0.05 eV above E$_F$
\cite{NatPhys}. More importantly, as the sample becomes thicker, the
gap (labeled by light blue shaded area in panels (a-c)) decreases
rapidly.  From single layer to trilayer graphene, the gap decreases from
0.26 eV to 0.14 eV then to 0.066 eV. For graphite, since the Dirac
point energy is above E$_F$ \cite{NatPhys}, whether there is a gap or
not cannot be directly addressed by ARPES.  However, from band
structure calculation, it is expected that the gap at the H point
is $\approx$ 0.008 eV \cite{McClure, DresselhausGIC}, which is
almost negligible.

\begin{figure}
\includegraphics[width=10 cm] {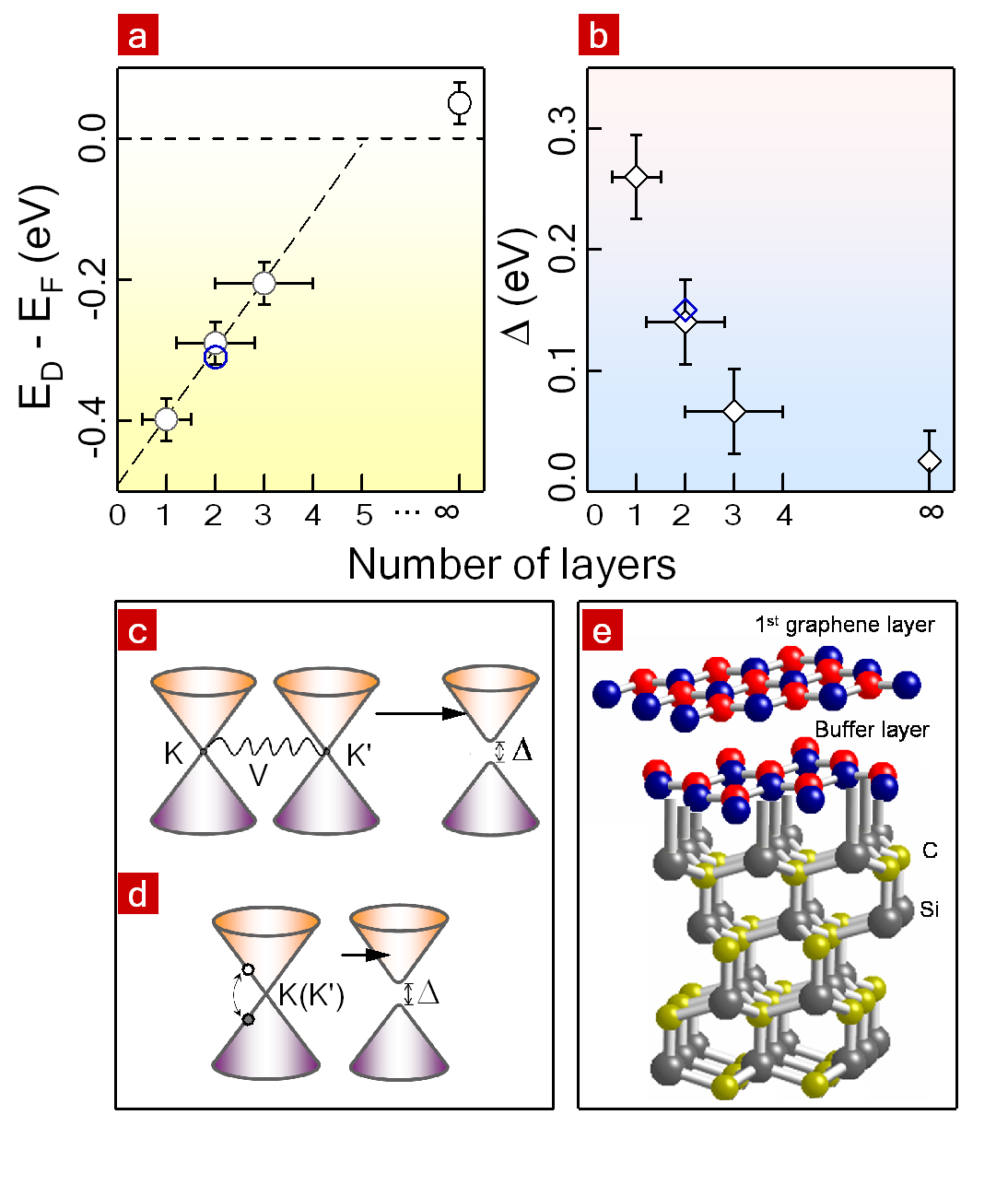}
\label{Figure 3} \caption{{\bf Thickness dependence of E$_D$ and $\Delta$.} (a,b) E$_D$ and $\Delta$ as
a function of sample thickness, for epitaxial graphene on 6H-SiC (black) and on 4H-SiC (blue).   The error bar for the sample thickness was taken from the XPS measurements \cite{Liz}.  For graphite, E$_D$ is extrapolated from the dispersions at k$_z$$\approx$$\pi$/c \cite{NatPhys}, and
the gap is estimated from band structure calculation \cite{McClure,
DresselhausGIC}.  (c, d) Two possible
mechanisms to open up a gap at the Dirac point. (e) Schematic drawing to show the inequivalent potentials on the A (blue) and B (red) sublattices induced by the interface.}
\end{figure}

Figures 3(a-b) summarize the evolution of the Dirac point energy
E$_D$ and the gap $\Delta$ for various sample thickness.  The layer
dependence of both quantities suggests that, beyond 5 layers,
epitaxial graphene behaves as bulk graphite \cite{NatPhys}.
The shift of E$_D$ in panel (a) is most likely due to the electric
field and surface charges present at the graphene-(n-SiC) interface \cite{BergerJPC}.  
We have also measured a bilayer graphene sample on a more insulating 4H-SiC substrate (Fig.2(b)) with resistivity of 10$^5$ $\Omega$/cm compared to 0.2 $\Omega$/cm in 6H-SiC.  In both cases, the Dirac point energy appears to be shifted by a similar amount below E$_F$, suggesting that the doping is most likely associated with the surface charges at the interface, rather than the carrier concentration of the substrate.  
This effect should decrease as the sample becomes
thicker, because the surface layer probed by ARPES is farther away
from the interface as the thickness increases. Also, the strong dependence of
E$_D$ on sample thickness is a direct manifestation of the short
interlayer screening length ($\approx$ 5 layers \cite{Paco}) of
graphene. This result shows that the sample thickness is an
effective way of controlling doping in epitaxial graphene.  Panel(b) shows the dependence of the gap on the sample thickness. A gap
in bilayer graphene has been reported and attributed to the
different potentials in the two graphene layers induced by doping or
electric field \cite{Castro, McCann, Eli}.  While this could
contribute to the gap in bilayer and even trilayer graphene, it
certainly is not the reason for the gap in the single layer graphene.

In the following, we discuss two possible scenarios and we propose that the gaps in single, double and triple
layer graphene are results of symmetry breaking due to the
substrate. As discussed in the introduction, there are two ways to
open up energy gaps at K and K$^\prime$. The scenario that invokes
the inter-Dirac-point hybridization (Fig.3c) requires translation
symmetry breaking.   The two known reconstructions on epitaxial
graphene, 6$\times$6 and ($6\sqrt3\times6\sqrt3$)R30$^{\circ}$
\cite{Tsai} are obvious candidates for the source of this
symmetry breaking. However, in order to mix K and K$^\prime$, a large
scattering wave vector is required. This is much longer than the
reciprocal lattice vectors of both reconstructions mentioned above.  High ordering process involving consecutive small
scattering wave vectors will be weak in general. Another source of inter-Dirac-point scattering is impurity scattering, which, as recently shown, can mix the wave functions at the two K points \cite{STM, Rutter}.  This however would give rise to a gap that strongly 
depends on the impurity concentration, in contrast to our finding. 
The gap is in fact the same in all the samples that we have studied, prepared 
under different conditions (with and without hydrogen etching of the SiC substrate)
and on differently doped substrates, insulating vs slightly electron doped substrate.

In our opinion, the more likely scenario is the breaking of the A, B sublattice symmetry.
This leads to the rehybridization of the valence and conduction band
states associated with the same Dirac point (Fig.3d), resulting in a gap at all the K and K$^\prime$ points. A necessary
prediction of this scenario is the breaking of the six fold rotational
symmetry of graphene near the Dirac point energy. 
For energy well above and/or below E$_D$, 
the symmetry is restored.
For bilayer and trilayer graphene, the breaking
of the A, B sublattice equivalence can be a direct consequence of the
the AB stacking between different layers. Indeed, topographic Scanning Tunneling Microscopy (STM) 
images for bilayer graphene have clearly shown inequivalent A and B
sublattices \cite{STM, Victor, First}, similar to what has been observed for graphite \cite{STMGraphite}. 
This simply derives from the fact that one sublattice has carbon atoms
directly below it while the other does not. Naively it seems that
this explanation will not work for single layer graphene.  However,
it is known that for epitaxially grown graphene, a buffer layer (Fig.3e) exists
\cite{EffectSubstrate, Seyller}. ARPES study of the buffer layer has
shown practically the same $\sigma$ bands as graphene while very
different $\pi$ bands \cite{Seyller}. This is because the $\pi$ orbitals have
hybridized with the dangling bonds from the substrate.
The fact that the $\sigma$ bands are unchanged suggests that, like
graphene, the carbon atoms in the buffer layer have also the
honeycomb arrangement with similar bond length.  Consequently, although the buffer layer is electronically inactive (absence of $\pi$ orbitals) \cite{Seyller}, structurally it can break the A, B sublattice symmetry when a single layer of graphene grows upon it (Fig.3e). This is particularly so in view of the small layer separation of $\approx$ 2 $\AA$ \cite{EffectSubstrate} and the AB stacking usually expected for very
thin graphene samples.  

For the single and bilayer graphene, we use a tight binding model with symmetry breaking on the A and B sublattices
to fit the symmetry breaking parameters 
to the observed energy gap (see supplementary information for the Bloch Hamiltonian). 
By fitting the dispersion, the symmetry breaking parameter in single layer graphene, defined as half of the difference between the substrate potentials on the A and B sublattices, is determined to be $m\approx 0.13$ eV.  In bilayer graphene, the symmetry breaking parameters in the top and bottom layers are $m_1\approx 0.49$ eV, $m_2\approx -0.21$ eV respectively. The magnitude of the symmetry
breaking parameter is much bigger in the bottom graphene layer than that in
single layer graphene, because it is sandwiched between the buffer layer 
and the top graphene layer. The reason for $m_2$ to have the opposite
sign is because of the AB stacking. This cancels part of the effect
in the bilayer graphene and decreases the gap. Therefore, for AB
stacking graphene, the eigen functions average out for many layers,
and the gap decreases rapidly.

\begin{figure}
\includegraphics[width=15 cm] {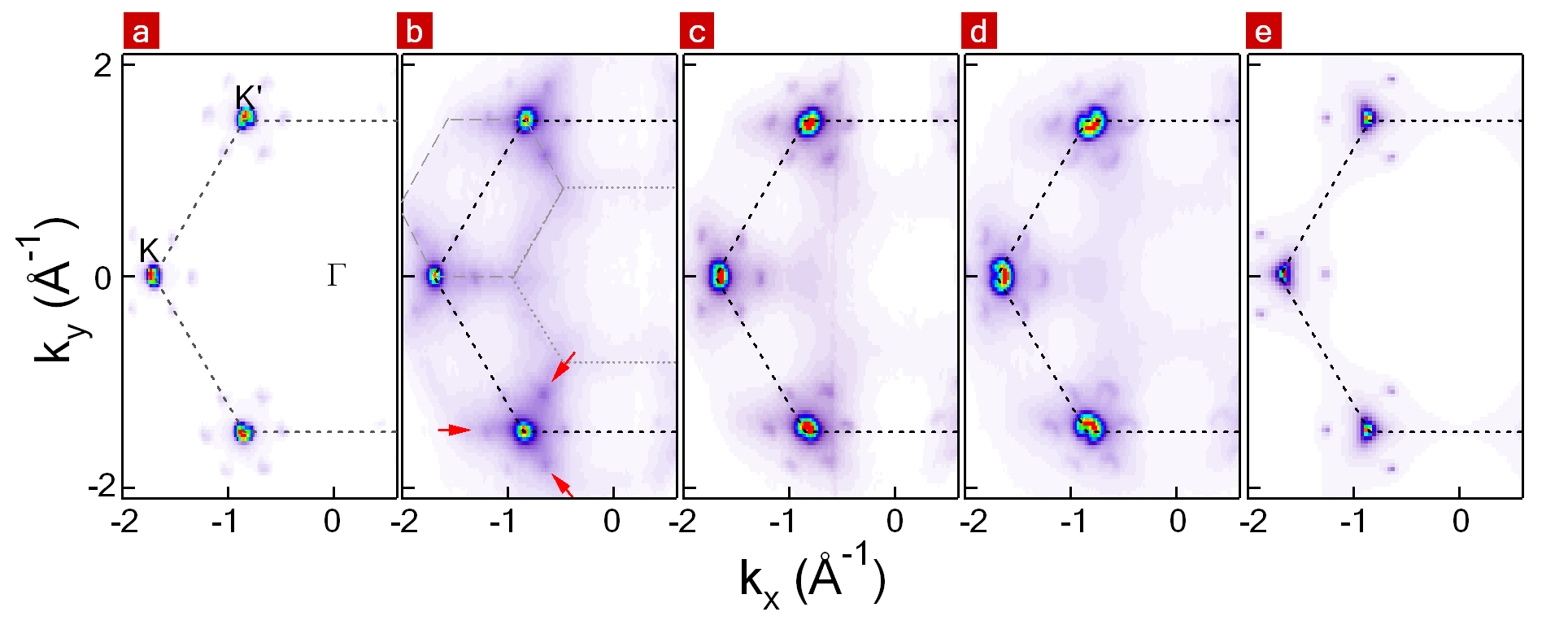}
\label{Figure 4} \caption{ {\bf Breaking of the six fold symmetry in the intensity map near E$_D$.} (a-d) ARPES intensity maps taken on single layer graphene at E$_F$, E$_D$,
-0.8 eV and -1.0 eV respectively. Near E$_D$ (panel b), the intensity of the six replicas near K shows breaking of six fold symmetry.  Note that to enhance the additional feature around
E$_D$, the color scale is saturated for the dominant features near K
and the replicas. (e) ARPES intensity map of the calculated spectral function at E$_D$ 
in the presence of symmetry breaking on the two carbon sublattices.}
\end{figure}

Figure 4 shows additional support for the A, B sublattice symmetry
breaking. Panels (a-d) show intensity maps taken on single layer
graphene as a function of k$_x$ and k$_y$ at E$_F$, -0.4, -0.8 eV
and -1.0 eV respectively.  The dominant features in these panels are
the small pockets centered at the six corners of the Brillouin zone.
Interestingly, around each corner, there are six faint replicas
forming a smaller hexagon. The intensity associated with them is
$\approx$ 4\% of the main intensity. Closer inspection shows that
the vectors connecting the center of the small hexagon to its six
corners are nearly the same as the second shortest reciprocal lattice vectors of the
($6\sqrt3\times6\sqrt3$)R30$^{\circ}$ \cite{BergerJPC, Liz} observed in low energy electron diffraction
LEED \cite{BergerJPC, Liz}. 
As E$_D$ is approached, three among the six faint replicas become more intense
(pointed by red arrows in Fig.4(b)). This suggests the breaking of the six
fold rotational symmetry of graphene down to three fold, and is
consistent with the notion of A and B sublattices being
inequivalent. 
In Fig.4(e), we use a tight binding model to compute the intensity of the replicas at E$_D$. 
The potential modulation imposed by the ($6\sqrt3\times6\sqrt3$)R30$^{\circ}$ reconstruction has been added as a perturbation to the Hamiltonian, and the sublattice symmetry
breaking has also been taken into account (see supplementary information). The result favorably agrees with the
observation.
We note that STM measurements on epitaxially grown single layer graphene 
do not show this symmetry breaking.  This is because the main graphene signal measured is near E$_F$, 
where no symmetry breaking is observed (see Fig. 4a).

In addition to these faint replicas, we observe additional intensity enhanced along the edge of certain medium sized hexagons around E$_D$ (see gray broken and dotted lines in panel b). The origin of this intensity is still
unclear. However, two observations can be made. 1) The center
mid-sized hexagon around $\Gamma$ (gray dotted lines) almost overlaps the first Brillouin zone of SiC.
2) All other hexagons (e.g. gray broken lines) are not regular, i.e. the six sides forming the hexagon do not have the same length.  Interestingly, they all pass through K and K$^\prime$.  Whether this reveals the presence
of perturbation that can hybridize the states at K and K$^\prime$
remains unclear.

In conclusion, we have reported the presence of an
energy gap at the K point in epitaxial graphene and we propose that it is induced by the interaction with the substrate. Thus if one can change the strength of the interaction by changing the substrate on which graphene is grown, a control of the gap size can be possibly achieved.  Since the epitaxial graphene is usually electron doped and the gap in this case is below E$_F$, the next important step to make graphene a semiconductor is to dope graphene with holes or to apply a gate voltage to move E$_F$ inside the gap region.  

{\bf Methods}

Atomically-thin graphene samples have been epitaxially grown by
thermal decomposition of a Si-terminated n-type SiC wafer at
increasing temperature \cite{BergerJPC, Liz}. The details of  the
growth process and characterization of surface quality using low
energy electron diffraction (LEED), and scanning electron microscopy
(SEM) have been discussed elsewhere \cite{Liz}.  The thickness of
the sample has been determined using Auger spectroscopy
\cite{BergerJPC} and X-ray photoemission spectroscopy (XPS)
\cite{Liz} as well as ARPES.  The absence of k$_z$ dispersion in ARPES over a large
momentum range of 4$\pi$/c (see Fig. S1 in supplementary information) confirms the thickness of the single
layer graphene.

ARPES data have been taken at Beamline 12.0.1 of the Advanced Light
Source (ALS) in Lawrence Berkeley National Lab with photon energy of
50 eV (Figs.1 and 2 except Fig.2d) and Beamline 7.0.1 with photon
energy of 140 eV (Fig.2d) and 100 eV (Fig.4).  The energy resolution
is 20-35 meV. The samples were measured at 25K with vacuum
better than 5.0$\times$10$^{-11}$ Torr.

\begin{acknowledgments}

We thank A. Geim and A.H. MacDonald for useful discussions, J. Graf for experimental assistance.  This work was supported by the National Science Foundation through Grant
No.~DMR03-49361, the Director, Office of Science, Office of Basic
Energy Sciences, Division of Materials Sciences and Engineering of
the U.S Department of Energy under Contract No.~DEAC03-76SF00098,
and by the Laboratory Directed Research and Development Program of
Lawrence Berkeley National Laboratory under the Department of Energy
Contract No. DE-AC02-05CH11231. A.~H.~C.~N was supported through NSF
grant DMR-0343790.  S. Y. Zhou thanks the Advanced Light Source Fellowship 
for financial support.

\end{acknowledgments}

Correspondence and requests for materials should be addressed to
Alessandra Lanzara (Alanzara@lbl.gov).

\end{document}